# Fluorescence blinking statistics from CdSe core and core-shell nanorods

*Siying Wang,[†] Claudia Querner,[†] Thomas Emmons,[‡] Marija Drndic,[*,†] and Catherine H. Crouch[*,‡]*

[†] Department of Physics and Astronomy, University of Pennsylvania, 209 South 33[rd] St., Philadelphia, PA 19104

[‡] Department of Physics and Astronomy, Swarthmore College, 500 College Ave., Swarthmore, PA 19081

Email: drndic@physics.upenn.edu, ccrouch1@swarthmore.edu

**RECEIVED DATE (to be automatically inserted after your manuscript is accepted if required according to the journal that you are submitting your paper to)**

TITLE RUNNING HEAD. CdSe nanorod blinking statistics.

[*] To whom correspondence should be addressed: Marija Drndic (phone: (215) 898-5810, fax: (215) 898-2010, e-mail: drndic@physics.upenn.edu), Catherine H. Crouch (phone: (610) 328-8386, fax: (610) 328-7895, e-mail: ccrouch1@swarthmore.edu).

**Abstract**

We report fluorescence blinking statistics measured from single CdSe nanorods (NRs) of seven different sizes with aspect ratio ranging from 3 to 11. This study included furthermore core/shell CdSe/ZnSe NRs and core NRs with two different surface ligands producing different degrees of surface passivation. We compare the findings for NRs to our measurements of blinking statistics from spherical CdSe core and CdSe/ZnS core/shell nanocrystals (NCs). We find that for both NRs and spherical NCs, the off-time probability distributions are well described by a power law, while the on-time probability distributions are best described by a truncated power law, $P(\tau_{on}) \sim \tau_{on}^{-\alpha} e^{-\tau_{on}/\tau_c}$. The measured crossover time $\tau_c$ is indistinguishable within experimental uncertainty for core and core-shell NRs, and for core NRs with different ligands, for a same core size, indicating that surface passivation does not affect the blinking statistics significantly. We find that at fixed excitation intensity, $1/\tau_c$ increases approximately linearly with increasing NR aspect ratio; for a given sample, $1/\tau_c$ increases very gradually with increasing excitation intensity. Examining $1/\tau_c$ vs. single-particle photon absorption rate for all samples indicates that the change in NR absorption cross-section with sample size can account for some but not all of the differences in crossover time. This suggests that the degree of quantum confinement may be partially responsible for the aspect ratio dependence of the crossover time.

**Keywords.**

## Introduction

Fluorescence intermittency, also called blinking, is a widely observed property of single fluorophores, from colloidally synthesized semiconductor nanocrystals (NCs) or quantum dots (QDs)[1] and polymer nanoparticles[2] to organic dye molecules[3] and green fluorescent protein (GFP).[4] Rather than steadily emitting light under continuous excitation, the fluorescence from individual NCs turns "on" and "off," with individual "on" or "off" events lasting from microseconds to many minutes.[1] This behavior has been widely studied in spherical CdSe NCs, both experimentally and theoretically,[5-32] as well as in some other materials such as PbS.[33] The random and uncontrollable nature of NCs blinking is a major obstacle to single-NC optoelectronic applications such as lasers[34] and single-photon sources,[21] as well as to using single NCs as biological fluorescent markers.[35,36] Consequently, understanding blinking may facilitate many single-NC applications.

Although the mechanism of blinking is not fully understood, it is thought that NCs become "dark" - cease emitting light - when one of the charge carriers in the photoexcited exciton becomes trapped at the surface of the NC, or tunnels entirely off the NC into the environment, leaving a net charge delocalized in the NC core. Fluorescence then resumes once the core regains electrical neutrality.[1,5,9,10] The exact mechanism is still under theoretical and experimental investigation.

Blinking in semiconductor NCs differs significantly from blinking observed from many other single fluorophores in the probability distribution of "on" or "off" events of a particular duration. If a histogram of the duration $\tau_{off}$ of all "off" events observed from each NC is calculated, the resulting probability distribution of the off times, $P(\tau_{off})$, has been found for spherical core/shell CdSe/ZnS NCs to follow a power law, $P(\tau_{off}) = A\tau_{off}^{-\alpha}$ ($\alpha \sim 1.5$).[7,10,14,17,19,20,24,28] The probability distribution of on-times, $P(\tau_{on})$, likewise follows a power law for short $\tau_{on}$, but falls below the power law at longer "on" times at room temperature. Probability distributions observed from most other single fluorophores are exponential or near-exponential.[27]

Recently, novel anisotropic NC shapes have been synthesized.[37] These are expected to offer a rich variety of electrical and optical properties; solar cells[38] and transistors[39] based on these shapes have been proposed. One example is the nanorod (NR), a NC that is elongated along the crystal *c*-axis so that charge carriers are most strongly quantum-confined along the NR cross-section. Studying the optical and electrical properties of NRs offers the possibility of observing the transition from quantum states confined in all three dimensions (0D states), as in a spherical NC, to states confined in only two dimensions (1D states). The elongation of NRs also makes them better conductors than spherical NCs, and thus more suitable for certain device applications.[38,40] While a variety of fluorescence measurements have been made on single NRs[40–45] and other elongated nanocrystals,[46,47] blinking statistics from NRs have not been reported to date.

In this paper, we report the first measurements of blinking statistics from single CdSe NRs. We study seven different types of NR, with lengths ranging from 18 to 38 nm and diameters from 3.5 to 6.9 nm, giving aspect ratios from 3.5 to 11.2. For four sizes of NRs, we studied both core CdSe and core/shell CdSe/ZnSe NRs; for one size we studied core NRs with two different surface ligands - trioctylphosphine oxide and hexadecylamine - which produce different degrees of surface passivation. We compare the measured probability distributions of off- and on-times to those measured from spherical NCs, both core and core-shell. This set of samples allows us to distinguish the effects of surface passivation and shape on the blinking statistics. We fit the on-time distributions to a truncated power law, $P(\tau_{on}) \sim \tau_{on}^{-\alpha} e^{-\tau_{on}/\tau_c}$. Finally, we investigate the effect of excitation intensity and absorption cross-section on the on- and off-time statistics to gain further insight into NR blinking.

**Experimental Methods**

**Synthesis of CdSe NRs.** We synthesized CdSe NRs of different lengths and diameters (Table 1), capped with a 1.1 nm layer of trioctylphosphine oxide (TOPO), by adapting literature procedures.[48-50] A ZnSe shell was grown on some of the core samples according to Ref. 51; hexadecylamine (HDA)-capped NRs (NR5-HDA) were prepared by heating TOPO-capped NRs in HDA for 1 h and precipitating/washing with methanol. Transmission electron microscopy (TEM, JEOL 2010F) confirmed the monodispersity (Figure 1a) and crystallinity (Figure 1b) of the NRs. Spherical CdSe and core/shell CdSe/ZnS NCs with core diameters of 5.2 nm were purchased from Evident Technologies. The samples studied are listed in Table 1.

**Measurement of the absorption cross-section $\sigma_\lambda$.** We prepared a series of concentrations of dispersions of NCs and NRs in toluene, including at least 7 different concentrations between ~$5 \cdot 10^{-8}$ and ~$5 \cdot 10^{-6}$ mol·l$^{-1}$ (maximum concentration of $6 \cdot 10^{-7}$ mol·l$^{-1}$ for NR6 and NR7). Absorption spectra were measured using a USB2000-VIS-NIR-spectrometer (Ocean Optics; integration time: 30 ms; resolution: 1 nm; path length: 1 cm). A Beer's Law plot of the absorbance intensities at the excitonic peak and at 532 nm (the laser wavelength used to excite fluorescence) was made to determine the molar extinction coefficients $\varepsilon_\lambda$. The per particle absorption cross-section $\sigma_\lambda$ was then calculated according to Ref. 52 (Table 1).

**Fluorescence Measurements.** We performed wide-field fluorescence imaging[53] of a very sparse sample of NCs or NRs, using an epifluorescence microscope (Olympus) with a 100× 0.95 NA dry objective. Samples were prepared by drop- or spin-casting a very dilute toluene solution of CdSe NCs or NRs onto a freshly cleaved mica substrate. The concentration of the solution was chosen so that individual NCs or NRs were typically separated by a few micrometers. The sample was illuminated by 532 nm light from a continuous-wave (cw) frequency-doubled YAG laser (Coherent Compass). The excitation intensity used for most measurements was 210 W·cm$^{-2}$. To study intensity dependence, the intensity varied from 90 to 1000 W·cm$^{-2}$. All samples were measured at room temperature in air immediately after preparation to minimize sample oxidation. Fluorescence movies were captured by a thermoelectrically cooled CCD camera (Princeton Instruments Cascade 512F) at 10 frames per second. All measurements presented in this paper lasted 2000 s unless otherwise specified. A background image measured from a clean mica substrate was subtracted from each frame of the movies. Individual emitters were identified in the image acquisition software and the fluorescence intensity $I(t)$ of each emitter was determined in each frame throughout the entire movie.

**<add Figure 1 here>**

**<add Table 1 here>**

**Results**

**Statistical Analysis of Fluorescence "On" and "Off" Times.** Figure 2a shows an example of the time-dependent fluorescence intensity, $I(t)$, measured for 2000 s from a single 5×18 nm NR (NR4). To define the threshold above which the NR is considered "on" we measured $I(t)$ in ten "dark" regions of the sample (*i.e.*, regions with no NRs) and found the greatest intensity range $\Delta I_{dark}$ and standard deviation $\sigma_{dark}$ represented among those ten. As shown in Figure 2a, the "on" threshold (solid line) for each NR is set by adding $\Delta I_{dark} + \sigma_{dark}$ to the minimum intensity measured from that NR (dotted line).

From $I(t)$ for a single NR, we determined the probability density of "off" or "on" events of duration $\tau_{off(on)}$. Probability density is commonly defined as

$$P\left(\tau_{off(on)}\right) = \frac{N\left(\tau_{off(on)}\right)}{N_{off(on)}^{tot}} \times \frac{1}{\Delta t}, \quad (1)$$

where $N\left(\tau_{off(on)}\right)$ is the number of "off" ("on") events of duration $\tau_{off(on)}$, $N_{off(on)}^{tot}$ is the total number of "off" ("on") events observed from that NR, and $\Delta t$ is the 100 ms frame duration of the movies. However, calculating the probability density from Eq. 1 assigns the same probability density to any $\tau_{off(on)}$ occurring only once during a particular experiment, and assigns a probability of zero to any $\tau_{off(on)}$ not observed in that experiment. A much longer experiment with more events would most likely give different probability densities for these rare events. We therefore calculated a weighted probability density according to the method of Kuno *et al.*[9]:

$$P\left(\tau_{off(on)}\right) = \frac{N\left(\tau_{off(on)}\right)}{N_{off(on)}^{tot}} \times \frac{1}{\Delta t_{off(on)}^{ave}}, \quad (2)$$

defining $\Delta t_{off(on)}^{ave} = (a+b)/2$, where $a$ and $b$ are the time differences to the next longest and next shortest observed event. For common event durations, $a$ and $b$ both equal the 100 ms frame duration and so $\Delta t_{off(on)}^{ave} = 100$ ms; $\Delta t_{off(on)}^{ave}$ increases for rare event durations if $a$ or $b$ exceeds 100 ms. This weighting scheme estimates the true probability of these rare events.

Previous studies[7,9,10] have used different approaches to fit the off-time probability density to a power law. We fit our measured off-time probability distributions for core-shell NCs (NCcs) with each approach (fits and discussion are provided in the Supporting Information). The same off-time data can give a power-law exponent ranging from 1.34 to 1.87 depending on fitting approach used. For all results in this paper, we binned our data by the 100 ms frame duration of the experiment and fit $P\left(\tau_{off}\right)$ to the power law $A\,\tau_{off}^{-\alpha}$ (rather than fitting a line to $\log\left[P\left(\tau_{off}\right)\right]$ vs. $\log\left[\tau_{off}\right]$). This approach gives somewhat (20-30%) lower exponents than the other approaches. We chose this approach because it minimizes manipulation of the data, and because the fits are dominated by the most reliable (short-duration) points in the probability distribution.

<add Figure 2 here>

<add Table 2 here>

**Off-Time Statistics.** The off-time probability density $P\left(\tau_{off}\right)$ obtained from the $I(t)$ data in Figure 2a is shown on a log-log scale in Figure 2b. We obtained similar distributions from each of 210 individual NRs of this sample (NR4). The probability density obtained by combining all events from all individual NRs observed (the "aggregated" probability density, $P\left(\tau_{off}\right)_{agg}$) is shown in Figure 2c. The probability densities obtained from individual NRs and from the aggregated data are well described by a power law. For the single NR shown in Figure 2b, the best-fit power law gives $\alpha_{off} = 1.30$; the distribution of exponents obtained from 210 individual NRs (inset to Figure 2b) has average value 1.3 and standard deviation 0.1. The power law fit to the aggregated results (Figure 2c) gives $\alpha_{off} = 1.22$, which is within one standard deviation of the average $\alpha_{off}$ obtained from individual NRs. We therefore estimate the uncertainty in $\alpha_{off}$ from the aggregated data to be 0.1, the standard deviation of the distribution, and conclude that the aggregated probability distribution is consistent with the range of individual probability distributions observed.

To investigate the effect of NR shape and surface passivation on blinking, we determined blinking statistics for all fourteen samples (listed in Tables 1 and 2). All aggregated off-time probability distributions show power law behavior, with the longest "off" times falling slightly below the power law. The distributions for the 5.2 nm-diameter core NC and NRs, *i.e.* NC, NR4, and NR5, are shown in Figure 2d.

The best-fit exponents of the power law fit to the aggregated data for all samples (Table 2) show no significant dependence of $\alpha_{off}$ on NR dimension, the presence of a ZnSe shell, or surface ligand; all values for NRs ($\alpha_{off} = 1.08$ to $1.22$) fall within the $\pm 0.1$ uncertainty range. The values obtained for the spherical NCs (core and core-shell) are slightly higher ($\alpha_{off} = 1.30$ and $1.34$).

**On-Time Statistics.** Although the off-time distributions for NRs and NCs are essentially indistinguishable and independent of shape and surface passivation, the on-time distributions show a distinct dependence on aspect ratio. Figure 3a shows the on-time probability density for the single NR4 data from Figure 2a; the aggregated probability density from all NR4 appears in Figure 3b. The shape of the on-time distributions obtained from each type of sample is similar to those shown in Figure 3b; the distributions follow a power law for on-times up to roughly 1 s for NRs and up to about 5 s for NCs, while longer "on" times fall below the power law, consistent with previous findings for NCs.[7-10]

To better match the shape of the on-time distributions, we fit them to a truncated power law,

$$P(\tau_{on}) = A \tau_{on}^{-\alpha_{on}} e^{-\tau_{on}/\tau_c} \qquad (3)$$

shown by the curves in Figure 3. This function can be used to describe a physical process which is governed by a power law at short times and an exponential at long times, as has been proposed for NC blinking;[23,29-31] the time $\tau_c$ then represents the crossover time between the two regimes. This function matches the shape of the on-time distributions well for both individual and aggregated data, and fits to this function consistently give $\chi^2$ one hundred times smaller than fits to a pure power law.

We also examined whether a stretched exponential probability distribution, $P(\tau_{on}) = A\exp(-\tau_{on}/\tau_c)^{-\beta}$, fits the on-time distributions (fits not shown), as these distributions are predicted from some models of disordered systems.[54] However, the $\chi^2$ for the fits was much poorer than that obtained with the truncated power law (comparable to that obtained with the pure power law). More significantly, the parameter $\tau_c$ varied by two orders of magnitude between individual NRs or NCs within the same sample, rather than displaying consistent values as with the truncated power law fit.

Comparing on-time distributions from different experiment durations (shown in the Supporting Information) indicates that increasing the experiment duration, so that more long-duration events are observed, produces an on-time distribution that increasingly resembles the distribution obtained by aggregating many shorter measurements. In addition, the histogram of individual crossover times calculated for the individual rods narrows, with the peak value staying the same within experimental uncertainty, and with the number of long-crossover-time outliers greatly reduced. We therefore conclude that aggregated distributions, if enough rods are included, serve as a reasonably good representation of the distribution that would be obtained from an extremely long experiment. As collecting very long measurements from single NRs or NCs is difficult, we therefore focus our analysis on aggregated data from many individual NRs, and use the distributions of crossover times for individual rods to determine the uncertainty in the value obtained from the aggregated data, as described in the supplementary materials.

The histogram of crossover times for the ensemble of NR4 is provided as an inset to Figure 3a. Aggregated results for the 5.2 nm diameter core samples (NC, NR4, and NR5) are provided in Figure 3c; Figure 4 shows on-time distributions from the three types of 5×28 NRs (NR5, NR5cs, and NR5-HDA). We obtain the shortest crossover time (0.89 s) for the longest NR (NR5); $\tau_c$ increases to 1.1 s for the shorter NR (NR4) and finally to 4.6 s for the NC. Equivalent data were obtained for all 14 samples in our study (Table 2). The exponents $\alpha_{on}$ are slightly smaller for NRs ($\alpha_{on} \sim 1.2$) than for NCs ($\alpha_{on} \sim 1.3$), but show no significant dependence on NR dimension or surface passivation, as shown in Table 2.

Surprisingly, we found the same on-time distribution and crossover time within experimental uncertainty for the core and core-shell 5×28 nm NRs (NR5 and NR5cs), as shown in Figure 4 and Table 2, in spite of the significant improvement in fluorescence quantum yield and resistance to photobleaching displayed by core-shell NRs. We also found very little difference between the core NRs capped with TOPO and with HDA, NR5 and NR5-HDA (Figure 4), in spite of an expected threefold difference in surface coverage. HDA is expected to attach to nearly all surface CdSe units while TOPO to only ~36%. This is because HDA has a linear geometry while TOPO has a cone-shaped geometry.[55]

Examining the data from all NC/NR samples studied, we find a single clear trend in crossover time: $\tau_c$ decreases fairly steadily with increasing NR aspect ratio. Quantitatively, it appears that $1/\tau_c$ increases approximately linearly with NR aspect ratio, as shown in Figure 5. The uncertainties in $\tau_c$ are determined from repeated measurements and from the distributions of $\tau_c$ values from the individual NRs, as described in the Supporting Information. We find that $\alpha_{on}$ varies somewhat more (~0.9-1.1) than $\alpha_{off}$, but with no clear trend.

Comparing the crossover times (Table 2) measured for 5.2 nm-diameter NRs to those for the 6.4 and 6.9 nm-diameter NRs, *i.e.* NR6 and NR7, we observe that for comparable length, a larger diameter reduces the crossover time. This is consistent with previous results from spherical NCs.[10] Unexpectedly, the 3.5 nm-diameter NRs (NR1-3) show smaller rather than larger crossover times, compared to the 5.2 nm-diameter NRs. The 3.5 nm-diameter NRs also bleached more rapidly than the larger NRs, requiring shorter data acquisition times which resulted in larger uncertainties in the values of $\tau_c$, as indicated in Table 2.

**<add Figure 3 here>**

**<add Figure 4 here>**

**<add Figure 5 here>**

**Effect of Excitation Intensity.** Finally, we examined the effect of changing the intensity of the exciting light on the NR off- and on-time statistics, by measuring blinking from 5×18 NRs (NR4) at eight different intensities from 90 W·cm$^{-2}$ to 1000 W·cm$^{-2}$. The off-time distribution is largely unaffected, as was found previously for spherical NCs[10]. Minor variations in the distribution tail produce a slight though fairly steady decrease with increasing intensity in $\alpha_{off}$ from 1.3 to 1.1 (Table 3). As this variation is within experimental uncertainty, it is not clear that this trend is significant.

The on-time distribution shows a clearer trend with intensity. Over the entire intensity range studied, the on-time distribution is well fit by a truncated power law. As intensity increases, $\tau_c$ decreases. The theory of Tang and Marcus predicts that $1/\tau_c$ should be proportional to the photon absorption rate,[29] and hence for a particular sample, proportional to the excitation intensity. As shown in Figure 6a, $1/\tau_c$ shows a small, steady increase with intensity. The exponent from the truncated power law, $\alpha_{on}$, also varies, but not in a consistent fashion, and only within uncertainty (Table 3).

The variation of $\tau_c$ with excitation intensity leads to the question: can the variation in $\tau_c$ with NR shape at fixed intensity be attributed entirely to the change in photon absorption rate due to changes in the single-particle absorption cross-section? Indeed, our measurements of the single-particle cross-section at the exciting wavelength, $\sigma_{532}$, for each sample (Table 1) indicate that the cross-section increases with increasing NR volume, as has been found for spherical NCs.[56] We therefore examined whether $\tau_c$ remains constant if the single-NR photon absorption rate is also held constant.

Figure 6b shows $1/\tau_c$ *vs.* the product of excitation intensity and single-particle cross-section (proportional to photon absorption rate) with points from all core samples measured. (core-shell samples have essentially the same $1/\tau_c$ as the core sample of the same size and thus would be redundant on this plot.) The $1/\tau_c$ values obtained from the larger-diameter NRs (NR5, NR6, and NR7) are fairly consistent with the measured intensity dependence of NR4, when intensity is converted to single-particle absorption rate. However, values obtained from the 3.5 nm-diameter NRs (NR1–3) and the spherical NCs do not map onto the measured NR4 intensity dependence.

**<add Figure 6 here>**

**<add Table 3 here>**

**Discussion**

Nanorod blinking displays many of the same features as blinking of spherical NCs: the off-time distributions follow a power law that is unaffected by excitation intensity or sample shape, and the on-time distributions follow a truncated power law with an intensity-dependent crossover time $\tau_c$. There are also significant differences between NR and NC blinking. In particular, $\tau_c$ is substantially shorter for NRs than for NCs, *i.e.*, NRs display far fewer long "on" events, and among a variety of NRs, $\tau_c$ decreases significantly with increasing NR aspect ratio.

Several important structural differences between NRs and NCs might be expected to contribute to these differences. First, because the NR surface is less sharply curved along its long dimension than the surface of spherical NCs (or the ends of NRs), and because the TOPO molecules typically used to cap both NRs and NCs are conical,[55] a lower percentage of surface CdSe units are attached to TOPO molecules on NRs than on NCs (Table 1), leading to less complete surface passivation. There should therefore be a larger number of non-passivated surface traps on NRs, which might be expected to reduce the lifetime of the "on" state. Second, the quantum confinement of the exciton in one-dimensional NRs is weaker than that in zero-dimensional spherical NCs. Consequently, there is less of a barrier than in spherical NCs for an excited charge carrier to tunnel to the surface or into the environment.[42,43] The greater likelihood of such tunneling events would likewise be expected to reduce the probability of long "on" times. Finally, the shorter crossover time in NRs might also be a consequence of the greater surface charge density on NRs.[57,58] Further experiments, such as comparing electrical force microscopy (EFM) measurements of surface charge and blinking statistics of individual NRs, could investigate this last point.

Our finding that the crossover time is the same within experimental uncertainty for core and core-shell NRs, and also for TOPO- and HDA-capped core NRs (Table 2 and Figure 4), indicates that surface passivation is not the primary determinant of NR blinking statistics. In core-shell NRs, the optically active core is almost perfectly passivated by the higher band gap semiconductor shell. Any defects arise from lattice mismatches between the two crystal structures; however, ZnSe and CdSe are very well lattice-matched.[59] In core NRs, the core surface is passivated only by the organic capping ligands, and the degree of passivation depends strongly on the type of ligand; the linear shape of the HDA molecule allows it to attach to nearly all CdSe surface units, while TOPO attaches to less than half of the surface units. The samples compared thus represent a wide range of surface passivation.

We did find, however, that core-shell spherical NCs had a somewhat greater crossover time than the corresponding core NCs (Table 2). One possible explanation for why the crossover time for NRs but not NCs is insensitive to surface passivation is that internal trap states, such as those induced by stacking faults in the crystal structure during growth, may affect carrier dynamics more than surface trap states in NRs. If so, improved surface passivation in NRs would not affect the crossover time substantially. Ultrafast measurements of carrier relaxation rates have likewise suggested that surface state trapping may be less significant in NRs than in NCs.[45]

Our observation of a very gradual increase of $1/\tau_c$ with excitation intensity (Figure 6a) is consistent with, though far less pronounced than, the intensity dependence of the on-time distribution observed for spherical NCs, which has previously been attributed to a reduced hopping rate at lower excitation intensities.[10] The intensity dependence we observe for the NR4, *i.e.* 5×18 nm NRs, is far weaker than that predicted by Tang and Marcus,[29] though it could be a linear relationship with a constant offset.

Examining measurements of $1/\tau_c$ from different samples as a function of photon absorption rate (Figure 6b), we find that changes in absorption cross-section can account for the changes of $1/\tau_c$ with length for the larger-diameter NRs (NR4–7), all of which have fairly similar aspect ratios (3.5 – 5.5). However, $1/\tau_c$ for the spherical NCs and the 3.5 nm-diameter NRs do not match that measured from the larger NRs with the same photon absorption rate. Our results therefore indicate that absorption cross-section changes cannot account for all of the observed variation in $1/\tau_c$ .

The roughly linear increase of $1/\tau_c$ with aspect ratio (Figure 5) could therefore be a manifestation of the reduced strength of quantum confinement with the increase in aspect ratio. If so, it also reveals that the transition from 0D to 1D confinement has a much greater effect on the crossover time than does decreasing confinement by increasing the diameter in spherical NCs. It could also reflect an increased number of internal trap states with increasing aspect ratio, or both effects could contribute. We postulate that changes in quantum confinement with NR aspect ratio may play a significant role and should be considered in future models.

**Conclusions**

In conclusion, we find that blinking statistics from CdSe NRs of a wide range of aspect ratios (3-11) display power-law off-time statistics and truncated power-law on-time statistics, with the crossover time for the on-time statistics decreasing with increasing aspect ratio or with increasing excitation intensity. We observe no significant difference in on-time statistics between TOPO-capped core, core-shell, and HDA-capped core NRs, indicating that surface passivation is largely unimportant in NR blinking, while we see a greater crossover time for core-shell than for core spherical NCs. We find that the variation in crossover time with aspect ratio for NRs can be partly but not completely explained in terms of changes in the absorption cross-section and hence the photon absorption rate.

We therefore attribute the shorter crossover time in higher aspect ratio rods to a combination of larger absorption cross-section, weaker quantum confinement, and possibly a higher incidence of internal trap states. In contrast to these differences in the on-time statistics, the off-time power law exponents do not depend on NR shape or surface coverage, and are very similar for NRs and NCs. Consequently, the mechanism determining the off-times is most likely the same for NRs and NCs, while the light-induced mechanism affecting the longer on-times sets in at shorter times in NRs than in NCs and is less sensitive to surface passivation. These findings indicate that blinking poses significant challenges for the use of single NRs in optoelectronic devices, and that the behavior of NR-based devices may be particularly sensitive to excitation intensity.

**Acknowledgements.** We thank Tara Finley and Nathan Landy for assistance with building the apparatus and initial experiments, Adam Cohen for assistance with preliminary data analysis, Michael D. Fischbein for performing transmission electron microscopy on NCs and NRs, and Hugo E. Romero for providing the 5×18 nm NR sample. This work was supported by the Howard Hughes Medical Institute (CHC), the NSF Career Award DMR-0449533 (MD), ONR awards YIP-N000140410489 and DURIP N00014-05-1-0393 (MD), and Swarthmore College.

**Supporting Information Available.** Details about different methods used to analyze the blinking data, the effect of experiment length on the blinking statistics, and the procedures used to determine the uncertainties in the fitting parameters. This material is available free of charge via the Internet at http://pubs.acs.org.

# References

(1) Nirmal, M.; Dabbousi, B. O.; Bawendi, M. G.; Macklin, J. J.; Trautman, J. K.; Harris, T. D.; Brus, L. E. *Nature* **1996,** *383*, 802.

(2) Vanden Bout, D. A.; Yip, W. T.; Hu, D. H.; Fu, D. K.; Swager, T. M.; Barbara, P. F. *Science* **1997,** *277*, 1074.

(3) Yip, W. T.; Hu, D. H.; Yu, J.; Vanden Bout, D. A.; Barbara, P. F. *J. Phys. Chem. A* **1998,** *102,* 7564.

(4) Dickson, R. M.; Cubitt, A. B.; Tsien, R. Y.; Moerner, W. E. *Nature* **1997,** *388*, 355.

(5) Efros, A. L.; Rosen, M. *Phys. Rev. Lett.* **1997,** *78*, 1110.

(6) Banin, U.; Bruchez, M.; Alivisatos, A. P.; Ha, T.; Weiss, S.; Chemla, D. S. *J. Chem. Phys.* **1999,** *110*, 1195.

(7) Kuno, M.; Fromm, D. P.; Hamann, H. F.; Gallagher, A.; Nesbitt, D. J. *J. Chem. Phys.* **2000,** *112*, 3117.

(8) Kuno, M.; Fromm, D. P.; Gallagher, A.; Nesbitt, D. J.; Micic, O. I.; Nozik, A. J. *Nano Lett.* **2001,** *1*, 557.

(9) Kuno, M.; Fromm, D. P.; Hamann, H. F.; Gallagher, A.; Nesbitt, D. J. *J. Chem. Phys.* **2001,** *115*, 1028.

(10) Shimizu, K. T.; Neuhauser, R. G.; Leatherdale, C. A.; Empedocles, S. A.; Woo, W. K.; Bawendi, M. G. *Phys. Rev. B* **2001,** *63*, 205316.

(11) Jung, Y.; Barkai, E.; Silbey, R. J. *Chem. Phys.* **2002,** *284*, 181.

(12) Verberk, R.; van Oijen., A. M.; Orrit, M. *Phys. Rev. B* **2002,** *66*, 233202.

(13) Ebenstein, Y.; Mokari, T.; Banin, U. *Appl. Phys. Lett.* **2002,** *80*, 4033.

(14) van Sark, W. G. J. H. M.; Frederix, P. L. T. M.; Bol, A. A.; Gerritsen, H. C.; Meijerink, A. *Chem. Phys. Chem.* **2002,** *3*, 871.

(15) Kuno, M.; Fromm, D. P.; Johnson, S. T.; Gallagher, A.; Nesbitt, D. J. *Phys. Rev. B* **2003,** *67*, 125304.

(16) Verberk, R.; Orrit, M. *J. Chem. Phys.* **2003,** *119*, 2214.

(17) Brokmann, X.; Hermier, J. P.; Messin, G.; Desbiolles, P.; Bouchaud, J. P.; Dahan, M. *Phys. Rev. Lett.* **2003,** *90*, 120601.

(18) Margolin, G.; Barkai, E. *J. Chem. Phys.* **2004,** *121*, 1566.

(19) Kobitski, A. Y.; Heyes, C. D.; Nienhaus, G. U. *Appl. Surf. Sci.* **2004,** *234*, 86.

(20) Hohng, S.; Ha, T. *J. Am. Chem. Soc.* **2004,** *126*, 1324.

(21) Brokmann, X.; Giacobino, E.; Dahan, M.; Hermier, J. P. *Appl. Phys. Lett.* **2004,** *85*, 712.

(22) Pelton, M.; Grier, D. G.; Guyot-Sionnest, P. *Appl. Phys. Lett.* **2004,** *85*, 819.

(23) Chung, I. H.; Bawendi, M. G. *Phys. Rev. B* **2004,** *70*, 165304.

(24) Muller, J.; Lupton, J. M.; Rogach, A. L.; Feldmann, J.; Talapin, D. V.; Weller, H. *Appl. Phys. Lett.* **2004,** *85*, 381.

(25) Issac, A.; von Borczyskowski, C.; Cichos, F. *Phys. Rev. B* **2005,** *71*, 161302.

(26) Margolin, G.; Protasenko, V.; Kuno, M.; Barkai, E. *arXiv:cond-mat* **2005** *v1, 0506512*.

(27) Frantsuzov, P. A.; Marcus, R. A. *Phys. Rev. B* **2005,** *72*, 155321.

(28) Yao, J.; Larson, D. R.; Vishwasrao, H. D.; Zipfel, W. R.; Webb, W. W. *Proc. Natl. Acad. Sci.* **2005,** *102*, 14284.

(29) Tang, J.; Marcus, R. A. *J. Chem. Phys.* **2005,** *123*, 054704.

(30) Tang, J.; Marcus, R. A. *Phys. Rev. Lett.* **2005,** *95*, 107401.

(31) Tang, J.; Marcus, R. A. *J. Chem. Phys.* **2005,** *123*, 204511.

(32) Chung, I.; Witkoskie, J. B.; Cao, J. S.; Bawendi, M. G. *Phys. Rev. E* **2006,** *73*, 011106.

(33) Peterson, J. J.; Krauss, T. D. *Nano Lett.* **2006,** *6*, 510.

(34) Chan, Y.; Caruge, J. M.; Snee, P. T.; Bawendi, M. G. *Appl. Phys. Lett.* **2004,** *85*, 2460.

(35) Bruchez, M.; Moronne, M.; Gin, P.; Weiss, S.; Alivisatos, A. P. *Science* **1998,** *281*, 2013.

(36) Dubertret, B.; Skourides, P.; Norris, D. J.; Noireaux, V.; Brivanlou, A. H.; Libchaber, A. *Science* **2002,** *298*, 1759.

(37) Manna, L.; Scher, E. C.; Alivisatos, A. P. *J. Am. Chem. Soc.* **2000,** *122*, 12700.

(38) Huynh, W. U.; Dittmer, J. J.; Alivisatos, A. P. *Science* **2002,** *295*, 2425.

(39) Cui, Y.; Banin, U.; Bjork, M. T.; Alivisatos, A. P. *Nano Lett.* **2005,** *5*, 1519.

(40) Millo, O.; Katz, D.; Steiner, D.; Rothenberg, E.; Mokari, T.; Kazes, M.; Banin, U. *Nanotechnology* **2004,** *15*, R1.

(41) Chen, X.; Nazzal, A.; Goorskey, D.; Xiao, M.; Peng, Z. A.; Peng, X. *Phys. Rev. B* **2001**, *64*, 245304.

(42) Rothenberg, E.; Ebenstein, Y.; Kazes, M.; Banin, U. *J. Phys. Chem. B* **2004**, *108*, 2797.

(43) Rothenberg, E.; Kazes, M.; Shaviv, E.; Banin, U. *Nano Lett.* **2005**, *5*, 1581.

(44) Le Thomas, N.; Herz, E.; Schops, O.; Woggon, U.; Artemyev, M. V. *Phys. Rev. Lett.* **2005**, *94*, 016803.

(45) Mohamed, M. B.; Burda, C.; El-Sayed, M. A. *Nano Lett.* **2001**, *1*, 589.

(46) Müller, J.; Lupton, J. M.; Lagoudakis, P. G.; Schindler, F.; Koeppe, R.; Rogach, A. L.; Feldmann, J.; Talapin, D. V.; Weller, H. *Nano Lett.* **2005**, *5*, 2044.

(47) Müller, J.; Lupton, J. M.; Rogach, A. L.; Feldmann, J.; Talapin, D. V.; Weller, H. *Phys. Rev. B* **2005**, *72*, 205339.

(48) Peng, Z. A.; Peng, X. G. *J. Am. Chem. Soc.* **2001,** *123*, 1389.

(49) Peng, Z. A.; Peng, X. G. *J. Am. Chem. Soc.* **2002,** *124*, 3343.

(50) Shieh, F.; Saunders, A. E.; Korgel, B. A. *J. Phys. Chem. B* **2005,** *109*, 8538.

(51) Reiss, P.; Bleuse, J.; Pron, A. *Nano Lett.* **2002**, *2*, 781.

(52) Leatherdale, C.A.; Woo, W.K.; Mikulec, F.V.; Bawendi, M.G. *J. Phys. Chem. B* **2002**, *106*, 7619.

(53) Moerner, W. E.; Fromm, D. P. *Rev. Sci. Instrum.* **2003,** *74*, 3597.

(54) Beadie, G.; Sauvain, E; Gomes, A. S. L.; Lawandy, N. M. *Phys. Rev. B* **1995**, *51*, 2180.

(55) Bullen, C.; Mulvaney, P. *Langmuir* **2006**, *22*, 3007.

(56) Klimov, V. *J. Phys. Chem. B* **2000**, *104*, 6112.

(57) Krauss, T.D.; Brus, L.E. *Phys. Rev. Lett.* **1999**, *83*, 4840.

(58) Krishnan, R.; Hahn, M.A.; Yu, Z.; Silcox, J.; Fauchet, P.M.; Krauss, T.D. *Phys. Rev. Lett.* **2004**, *92*, 216803.

(59) Reiss, P.; Carayon, S.; Bleuse, J.; Pron, A. *Synth. Met.* **2003**, *139*, 649.

**Tables.**

**Table 1.** Characteristics of CdSe core spherical nanocrystal (NC) and nanorod (NR) samples used in this study.[a]

| Sample | $d$ (nm)[b] | $l$ (nm)[b] | $l/d$[b] | $A$ (nm$^2$)[c] | $V$ (nm$^3$)[c] | $N_{TOPO}/N_{CdSe(surf)}$ (%)[d] | $\sigma_\lambda$ ($10^{-15}$ cm$^2$)[e] | $\sigma_{532}$ ($10^{-15}$ cm$^2$)[e] |
|---|---|---|---|---|---|---|---|---|
| NC | 5.2 | - | 1 | 85 | 74 | 49 | 2.7 | 2.6 |
| NR1 | 3.4 | 18 | 5.3 | 192 | 153 | 45 | 5.3 | 4.0 |
| NR2 | 3.5 | 25 | 7.1 | 275 | 229 | 42 | 6.7 | 5.0 |
| NR3 | 3.4 | 38 | 11.2 | 406 | 335 | 42 | 7.9 | 6.1 |
| NR4 | 5.2 | 18 | 3.5 | 295 | 345 | 37 | 8.0 | 8.9 |
| NR5 | 5.2 | 28 | 5.4 | 457 | 558 | 36 | 8.3 | 9.6 |
| NR6 | 6.4 | 22 | 3.5 | 442 | 639 | 34 | 9.6 | 19 |
| NR7 | 6.9 | 34 | 4.9 | 737 | 1185 | 32 | 14 | 30 |

[a] Core-shell NC and NR samples are named according the core sample on which the shell was grown followed by "cs", e.g. NR1cs. [b] Diameter ($d$), length ($l$), and aspect ratio ($l/d$) determined from the absorption spectra and TEM analysis. [c] Surface ($A$) and volume ($V$) estimated from the dimensions of the particles assuming perfect sphere or rod shape. [d] Surface coverage of available CdSe surface sites with TOPO molecules estimated from the ligand volume according to the procedure described by Bullen and Mulvaney.[55] [e] Absorption cross-section at the excitonic peak ($\sigma_\lambda$) and at 532 nm ($\sigma_{532}$).

**Table 2.** Off-time (on-time) exponents $\alpha_{off(on)}$ and on-time crossover times $\tau_c$ obtained for CdSe core and core-shell NCs and NRs.[a]

| Sample | $\alpha_{off}$[b] | $\alpha_{on}$[c] | $\tau_c$ (s)[c] |
|--------|---------|--------|-----------|
| NC | 1.30 | 1.32 | 4.6 |
| NR1 | 1.17 | 1.18 | 0.60 [d] |
| NR2 | 1.08 | 0.96 | 0.36 [e] |
| NR3 | 1.16 | 0.98 | 0.44 [d] |
| NR4 | 1.22 | 1.10 | 1.1 |
| NR5 | 1.17 | 1.17 | 0.88 |
| NR6 | 1.23 | 1.05 | 1.0 |
| NR7 | 1.20 | 0.93 | 0.62 |
| NCcs | 1.34 | 1.35 | 7.1 |
| NR1cs | 1.18 | 1.12 | 0.66 |
| NR2cs | 1.17 | 1.14 | 0.50 |
| NR3cs | 1.17 | 1.2 | 0.59 |
| NR5cs | 1.22 | 1.10 | 0.95 [f] |
| NR5-HDA | 1.22 | 1.02 | 0.66 |

[a] The fits were performed on aggregated data of 100 nanoparticles. Except when otherwise specified, data were obtained from movies 2000 s long. [b] Off-time exponent obtained from power law fit (Eq. 2). [c] On-time fitting parameters obtained from truncated power law fit (Eq. 3). The average uncertainty in the values of $\tau_c$ was 20%. [d] 1200 s long movie (uncertainty +30%/-20%). [e] 600 s long movie (uncertainty +40%/-30%). [f] Mean value of two independent measurements. Determination of uncertainties is discussed in the supplementary materials.

**Table 3.** Intensity dependence of off- and on-time parameters for sample NR4.[a]

| $I$ (W·cm$^{-2}$)[b] | $\alpha_{off}$[c] | $\alpha_{on}$[d] | $\tau_c$ (s)[d] |
|:---:|:---:|:---:|:---:|
| 90 | 1.27 | 1.09 | 1.3 |
| 210 | 1.23 | 1.10 | 1.1 |
| 300 | 1.29 | 1.01 | 1.1 |
| 400 | 1.24 | 0.92 | 0.79 |
| 500 | 1.22 | 0.79 | 0.73 |
| 600 | 1.21 | 0.79 | 0.69 |
| 690 | 1.14 | 1.16 | 0.69 |
| 870 | 1.12 | 1.20 | 0.65 |
| 1000 | 1.08 | 1.04 | 0.40 |

[a] The fits were performed on aggregated data from 50 individual NRs. [b] Excitation intensity. [c] Off-time exponent obtained from power law fit (Eq. 2). [d] On-time fitting parameters obtained from truncated power law fit (Eq. 3).

**Figure Captions.**

**Figure 1.** (*a*) Low-resolution and (*b*) high-resolution transmission electron microscope (TEM) images of NR4 deposited on thin films of amorphous carbon supported by a copper grid.

**Figure 2.** (*a*) Representative intensity *vs*. time data $I(t)$ obtained from a single NR (sample NR4) with the threshold above which it is considered to be "on" indicated by the solid line. (*b*) Off-time probability density $P(\tau_{off})$ obtained from the data shown in (*a*). Inset in (*b*): Histogram of exponents $\alpha_{off}$ for best-fit power law for each of 210 individual NRs observed in this sample. (*c*) Aggregated probability density of off-times $P(\tau_{off})_{agg}$ obtained by combining all off-times from all individual NRs observed. (*d*) Comparison of $P(\tau_{off})_{agg}$ obtained from NC (green upright triangles), NR4 (black squares), and NR5 (red inverted triangles), offset vertically by multiples of three decades. The parameters of the best-fit power law (solid lines) are provided in Table 2.

**Figure 3.** (*a*) On-time probability density $P(\tau_{on})$ obtained from the data shown in Figure 2a. Inset: Histogram of crossover times $\tau_c$ for best-fit truncated power law for each of 210 individual NRs observed (sample NR4). (*b*) Aggregated on-time probability density $P(\tau_{on})_{agg}$ obtained by combining on-times from all individual NRs observed. (*c*) Comparison of $P(\tau_{on})_{agg}$ obtained from NC (green upright triangles), NR4 (black squares), and NR5 (red inverted triangles). The parameters of the best-fit truncated power law (solid lines) are cataloged in Table 2.

**Figure 4.** Aggregated on-time probability distributions for NR5 (red inverted triangles), NR5cs (orange squares) and NR5-HDA (blue diamonds), obtained from 100 NRs for each sample at 210 W·cm$^{-2}$. Best-fit exponents and crossover times are provided in Table 2.

**Figure 5.** Inverse crossover time *vs.* aspect ratio for all TOPO-capped core NC and NR samples studied, along with best-fit line, $1/\tau_c = 0.23 + 0.24 \times$ aspect ratio.

**Figure 6.** (a) Excitation intensity dependence of the inverse crossover time of NR4. Best-fit exponents and crossover times are provided in Table 3. (b) Inverse crossover times of measured core NRs and NC as a function of excitation intensity $\times$ single-particle absorption cross-section (this quantity is proportional to the single-particle photon absorption rate).

**Figures.**

**Figure 1**

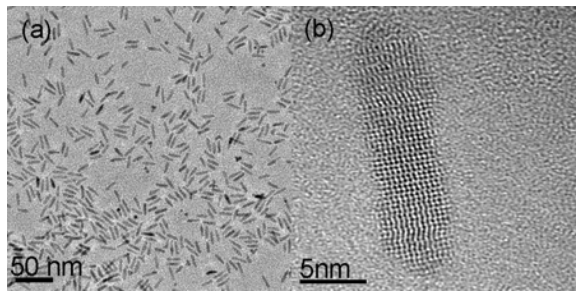

**Figure 2**

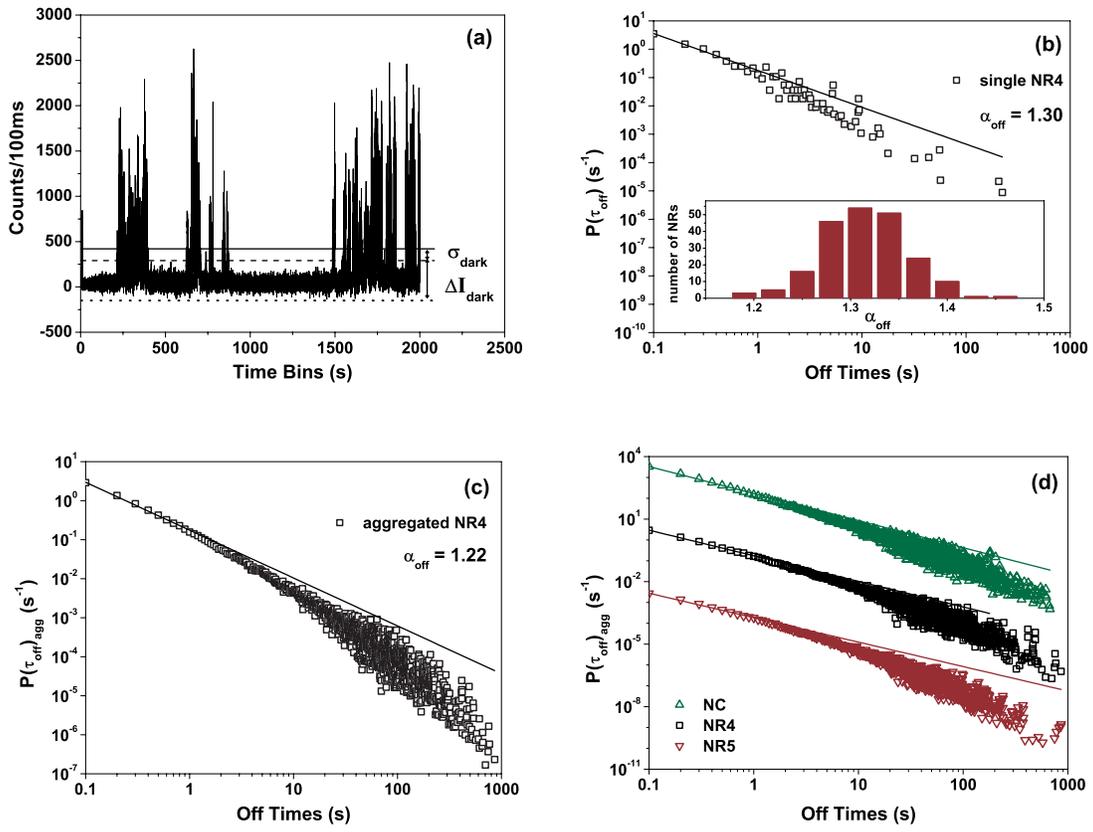

**Figure 3**

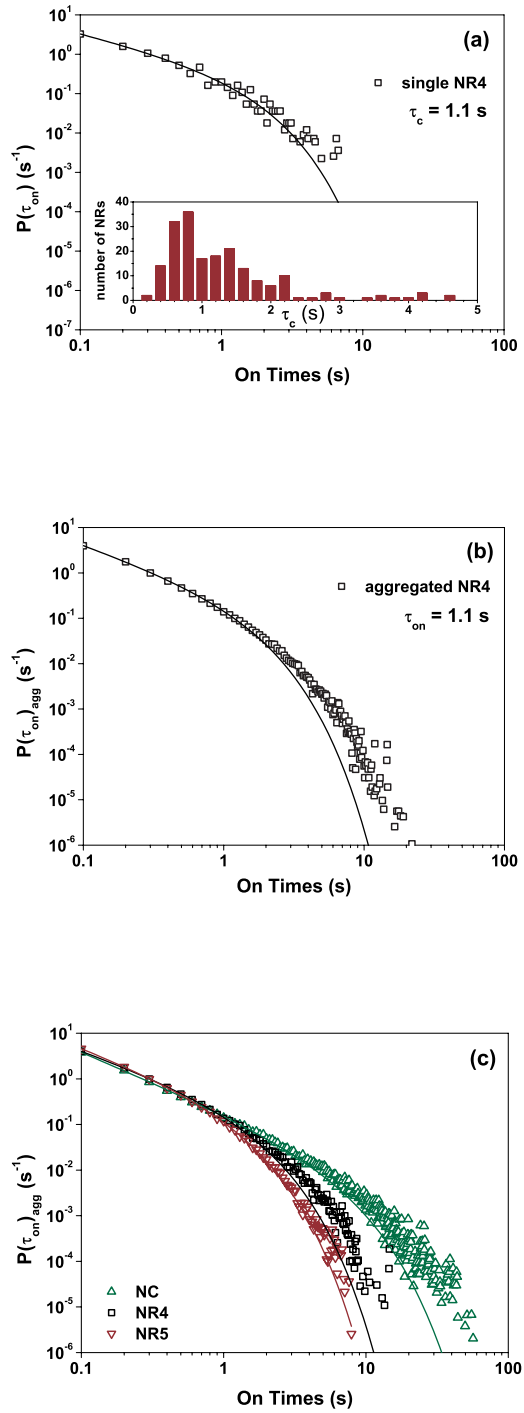

**Figure 4**

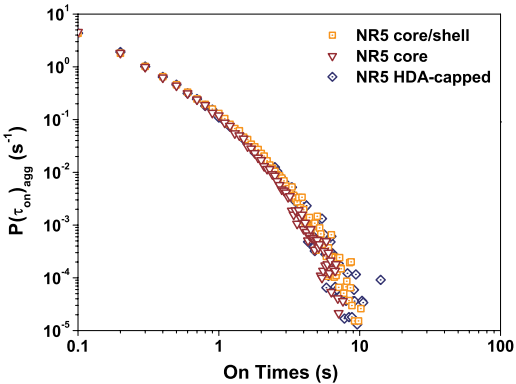

**Figure 5**

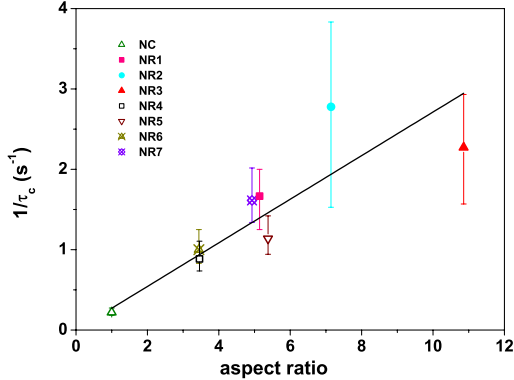

**Figure 6**

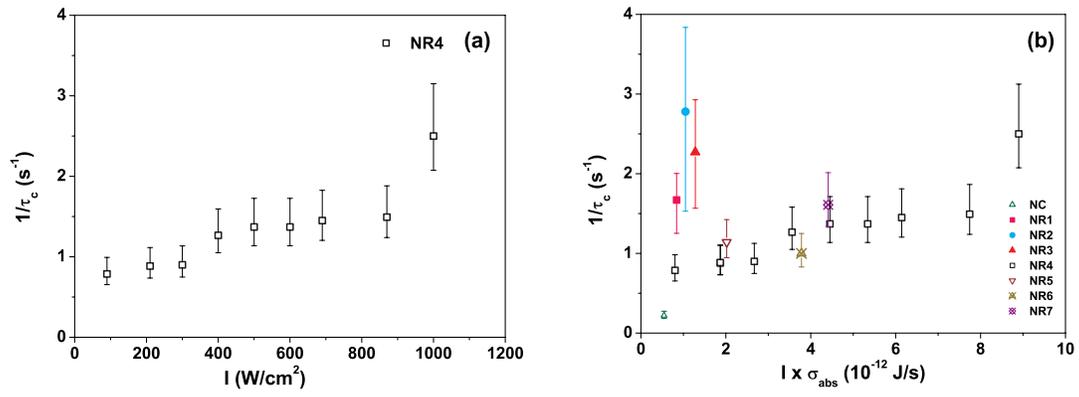

# Fluorescence blinking statistics from CdSe core and core-shell nanorods

*Siying Wang,[†] Claudia Querner,[†] Thomas Emmons,[‡] Marija Drndic,[\*,†] and Catherine H. Crouch[\*,‡]*

[†] Department of Physics and Astronomy, University of Pennsylvania, 209 South 33rd St., Philadelphia, PA 19104

[‡] Department of Physics and Astronomy, Swarthmore College, 500 College Ave., Swarthmore, PA 19081

## 1    Effect of data analysis methods on the off-time power-law exponent

Previous studies of blinking in core-shell NCs[1-3] have used a variety of approaches to fit the off-time probability density to a power law. In this section, we argue that the power law exponent depends significantly on the fitting approach and that the variation of exponents in the literature is probably partly due to how the raw data was analyzed. To determine the sensitivity of the exponent to the choice of fitting approaches, and to determine which approach best represents the data and minimizes the errors, we fit our measured off-time probability distributions for 5.2 nm-diameter core-shell NCs (NCcs — see Table 1 in article) — essentially the same diameter used in previous studies — using each approach.

There are two separate steps involved: calculation of the probability density and fitting to the power law. In calculating the probability density, the method involving the least data manipulation is to simply use the definition of probability density, $P\left(\tau_{off(on)}\right) = \dfrac{N\left(\tau_{off(on)}\right)}{N_{off(on)}^{tot}} \times \dfrac{1}{\Delta t}$ (Eq. 1) with $\Delta t$ the time resolution of the intensity *vs.* time data. Subsequently fitting this off-time density to the power law $P\left(\tau_{off}\right) = A\,\tau_{off}^{-\alpha_{off}}$, as shown in Figure S1a, gives $\alpha_{off} = 1.34$.

As can be seen in Figure S1a, the distribution calculated in this fashion reaches a plateau at long times, which is an artifact of the experiment duration; the probability of rare events occurring only a few times or not at all during a particular experiment is not the true statistical probability of those events. To address this shortcoming, Kuno *et al.*[2] introduced the weighting procedure described in the article, in which $\Delta t$ is replaced by $\Delta t_{off(on)}^{ave} = \left(a+b\right)/2$, where $a$ and $b$ are the time differences to the next longest and next shortest observed event. This affects only rare events whose duration is adjacent to durations with no observed events. Figure S1b shows the probability density obtained using this weighting scheme. Fitting this weighted probability density to a power law, we obtain the same exponent, $\alpha_{off} = 1.34$ (solid line in Figure S1b), that was obtained with the unweighted probability distribution. It is not surprising that the exponent is unchanged because the power-law fit is dominated by high-probability, short-duration points.

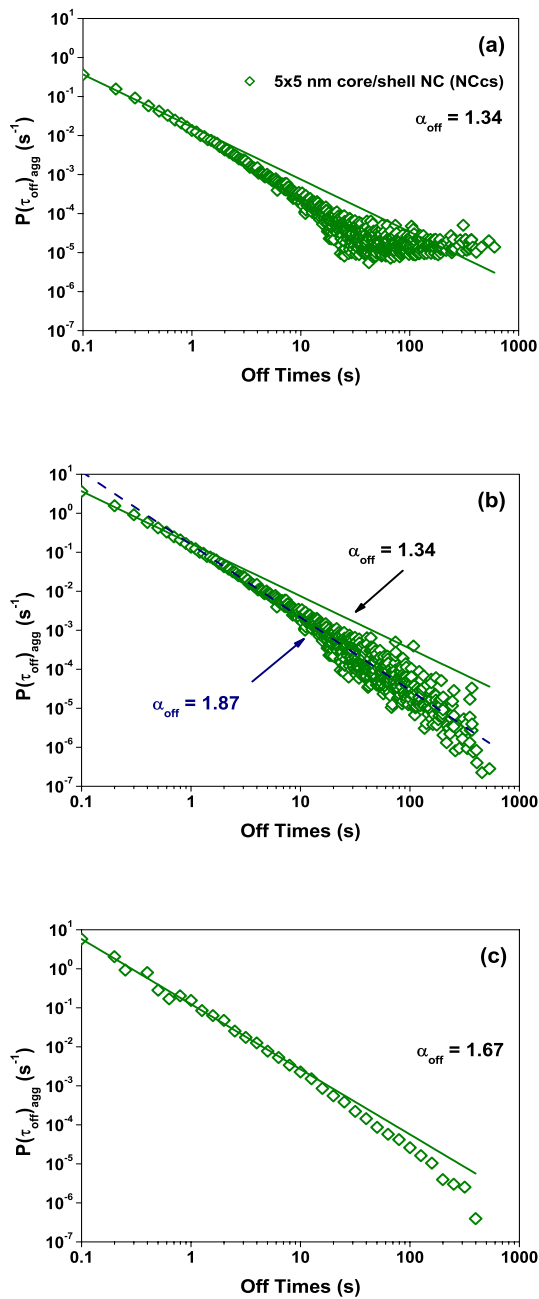

**Figure S1.** $P(\tau_{off})_{agg}$ obtained from 100 single NCs. Solid lines indicate the best-fit power law. (*a*) Unweighted distribution according to Eq. 1. (*b*) Distribution weighting the rare events according to Ref. 2. Dashed line represents the linear best-fit of the log-log-plot. (*c*) Distribution binned logarithmically.[3]

To further smooth the distribution, Shimizu *et al.* calculated the probability density from their data with only ten time bins in each decade of event duration; the length of the bins thus increases logarithmically.[3] Binning our data in this manner greatly reduces the scatter (Figure S1c), and increases the best-fit power-law exponent to $\alpha_{off} = 1.67$.

As seen in Figure S1b, the low-probability, long-duration points in the off-time distribution fall slightly below the power law (though not enough for a truncated power law fit to be successful). In their analysis, Kuno *et al.*[1,2] calculated $\log\left[P\left(\tau_{off}\right)\right]$ and $\log\left[\tau_{off}\right]$, and fit a line to the result. In such a fit (dashed line in Figure S1b), the longer-time events influence the fit as much as the shorter-time events, so that the best-fit line is steeper ($\alpha_{off} = 1.87$ for our data), with the shortest-time points falling below the fit.

We chose to bin our data by the 100 ms frame duration of the experiment, rather than using logarithmically increasing bins, to maximally preserve the measured results. We weighted the rare event probabilities according to Kuno's method, which does not alter the resulting exponents. We also performed power-law fits to $P\left(\tau_{off}\right)$ rather than linear fits to $\log\left[P\left(\tau_{off}\right)\right]$, as the power-law fit is dominated by the short time events which represent the most reliable part of the probability distribution. These choices result in our off-time exponents being somewhat smaller than would be obtained with either of the other two published approaches; the higher values obtained for our spherical NC data shown in Fig S1 are consistent with the values reported by other groups in the literature.

## 2     Effect of experiment length on individual NR on-time distributions

We compared on-time distributions from 2000 s and 4000 s movies taken of 5×18 nm NRs (NR4), to determine the reliability of using aggregated data to determine crossover times. Figure S2a shows that the longer measurements include more long-duration events and produce an on-time distribution that increasingly resembles the distribution obtained by aggregating many shorter measurements. Figure S2b shows that the histogram of individual crossover times calculated for the individual rods narrows as the experiment lengthens, while the peak value stays essentially the same within experimental uncertainty. In addition, when fitting individual NRs to the truncated power law, out of 100 NRs observed for 2000 s, 5 gave crossover times greater than 4.5 s and the fit failed to converge for 2 NRs. For 100 NRs observed for 4000 s, all fits converged and all gave crossover times less than 4 s, indicating that as the experiment time lengthens, those NRs which might have appeared at shorter times to be very different from the distribution become more similar.

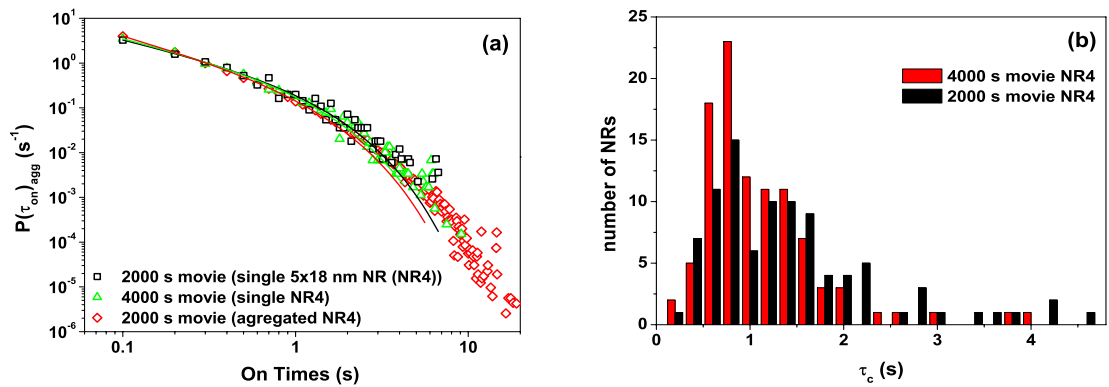

**Figure S2.** (a) On-time probability distributions obtained from a single 5×18 nm NR (NR4) measured for 2000 s (black) or 4000 s (green), and aggregated from 210 NRs measured for 2000 s (red). (b) Histograms of crossover times $\tau_c$ obtained from 100 NRs measured for 2000 s (black) or for 4000 s (red). Fits to the probability distributions for five of the NRs measured for 2000 s gave $\tau_c > 4.5$ s, which do not appear on the graph above, and the fits to two did not converge, while all of those measured for 4000 s gave fits that converged with $\tau_c < 4$ s.

## 3    Determination of uncertainties in values of $\tau_c$

We estimated the uncertainty in $\tau_c$ using several different approaches. Because the $\tau_c$ values provided in Tables 2 and 3 are determined from data aggregated from an ensemble of NRs or NCs, the uncertainty comes from the limited size of the sample and the variation between individual NRs in the sample. Two independent measurements of $\tau_c$ for the 5×28 nm core-shell NR (NR5cs), each determined from a fit to aggregated data from 100 single NRs, give 0.88 s and 1.02 s. Figure S3a shows the histogram of $\tau_c$ values found by fitting the on-time probability distributions for the individual NR5cs in the data set that gave the value of 0.88 s; the mean is 0.97 s (the standard deviation is 0.6 s). Considering both the two independent aggregated measurements and the histogram of individual measurements suggests that the uncertainty in the value from the aggregated dataset is approximately 0.1, or 10%. This value is consistent with the range of values found by selecting 100 NRs randomly out of the ensemble of 210 NR4 and aggregating those data to find $\tau_c$; repeating this process 200 times gave $\tau_c$ values between 0.99 s and 1.30 s. We therefore attribute an uncertainty of 20% to the measurements made by aggregating data from 100 NRs measured for 2000 seconds.

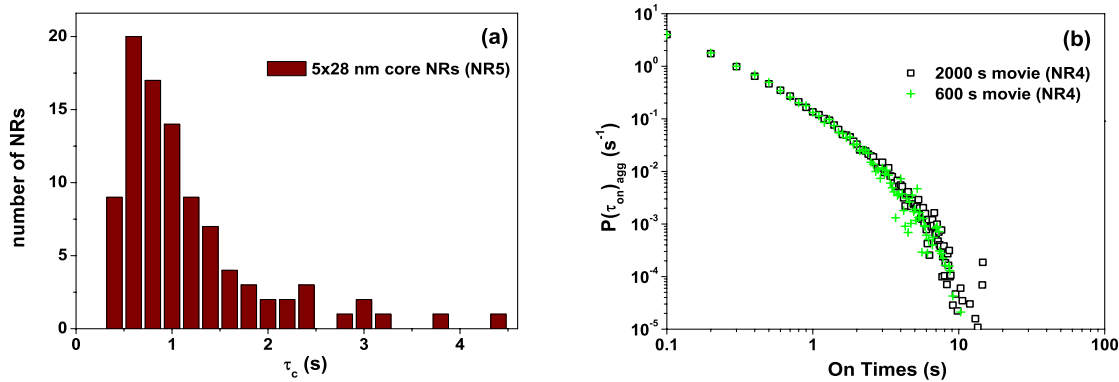

**Figure S3.** (a) Histogram of truncation times obtained for NR5. (b) On-time distributions obtained from (black) 2000 s movie and (green) 600 s excerpt from same 2000 s movie of sample NR4.

For the 3.5 nm diameter NRs, we measured fluorescence for only 1200 s or (for the 3.5×25 nm NR, *i.e.* NR2) 600 s due to rapid photobleaching of the sample. Shortening the experiments correspondingly increased the uncertainty. To estimate this uncertainty, we analyzed a shorter excerpt from a 2000 s movie of sample NR4. Figure S3 shows the on-time distributions obtained from the 2000 s full experiment and the 600 s excerpt; for the shorter dataset, the overall distribution is noisier at long times (though nearly identical for short times), and $\tau_c$ decreases by 20%. We therefore used error bars of +40% and –30%, with the larger positive error bar indicating the likelihood that the value of $\tau_c$ was underestimated for 600 s experiments. Following the same procedure with a 1200 s excerpt gives us an uncertainty estimate of +30% and –20% for the 1200 s movies.

### References

(1) Kuno, M.; Fromm, D. P.; Hamann, H. F.; Gallagher, A.; Nesbitt, D. J. *J. Chem. Phys.* **2000,** *112*, 3117.

(2) Kuno, M.; Fromm, D. P.; Hamann, H. F.; Gallagher, A.; Nesbitt, D. J. *J. Chem. Phys.* **2001,** *115*, 1028.

(3) Shimizu, K. T.; Neuhauser, R. G.; Leatherdale, C. A.; Empedocles, S. A.; Woo, W. K.; Bawendi, M. G. *Phys. Rev. B* **2001,** *63*, 205316.